# Seasonality Effects on Consumers' Preferences Over Quality Attributes of Different Beef Products

**Ali Ardeshiri [1*], Spring Sampson[2], Joffre Swait[3],**

1- Research Centre for Integrated Transport Innovation (rCITI), School of Civil and Environmental Engineering, University of New South Wales (UNSW) Sydney, NSW 2052 Australia.

2- Harvest Insights, South Melbourne, VIC 3205, Australia

3- JDS Behavior Insights, Taylors, SC 29687, USA

[*]- Corresponding author, Email: A.Ardeshiri@unsw.edu.au

## Abstract

Using discrete choice modelling, the study investigates 946 American consumers' willingness-to-pay and preferences for diverse beef products. A novel experiment was used to elicit the number of beef products that each consumer would purchase. The range of products explored in this study included ground, diced, roast, and six cuts of steaks (sirloin, tenderloin, flank, flap, New York and cowboy/rib-eye).

The outcome of the study suggests that US consumers vary in their preferences for beef products by season. The presence of a USDA certification logo is by far the most important factor affecting consumer's willingness to pay for all beef cuts, which is also heavily dependent on season. In relation to packaging, US consumers have mixed preference for different beef products by season. The results from a scaled adjusted ordered logit model showed that after price, safety-related attributes such as certification logos, types of packaging, and antibiotic free and organic products are a stronger influence on American consumers' choice.

Furthermore, US consumers on average purchase diced and roast products more often in winter "slow cooking season", than in summer; whereas New York strip and flank steak are more popular in the summer "grilling season".

This study provides valuable insights for businesses as well as policymakers to make inform decisions while considering how consumers relatively value among different labelling and product attributes by season and better address any ethical, safety and aesthetic concerns that consumers might have.

1. **INTRODUCTION**

In 2016, the United States (US) was Australia's second largest export destination. With 24% of Australia's beef exports and a value of AUS$1.7 billion, the US beef market plays an important role in Australia's export economy (Meat & Livestock Australia, 2017). A better understanding of American demand for beef is important as Australia faces strong competition from Canada and New Zealand in the US beef market. In 2016, it was reported that Australia was ranked number one in exporting beef and veal to the US. However, in the first-half 2017, Australia has exported relatively less beef than its competitors and the exported carcass weight has declined by 34% compared to the first half of 2016. A better understanding of US consumers' preferences is crucial if Australia is to maintain their position in the US beef import market. This study explores US consumers decisions over series of repeated hypothetical scenarios and evaluates how their value judgment for beef products are formed and may differ by season.

Beef demand, in its simplest form, is influenced by price and the ongoing evolving consumer preferences for taste, health benefits, convenience, etc. Preference for different cuts of beef are not only based on intrinsic and extrinsic cues, but also the context in which it is eaten.

One context that has been recognized as important by the econometricians is modelling seasonality effects on customer demand for goods and services (Lusk, Marsh, Schroeder, & Fox, 2001; Moskowitz & Beckley, 2009). Studies have looked at the importance of seasonality as a factor in beef purchasing habits, such as the seasonality effect on beef price (Capps, Farris, Byrne, Namken, & Lambert, 1994), quality grade cues (Farris & Holloway, 1990; Hogan Jr & Ward, 2003; Lusk et al., 2001), and hedging wholesale beef cuts (Namken, Farris, & Capps Jr, 1994; Schroeder & Yang, 2001). However, there has been no systematic analysis of the effect seasonality has on purchasing behaviour and preference for different beef cuts. Beef industries can benefit from understanding the seasonality effects on consumers decision and reduce the high failure rate of their new developed products (Dijksterhuis, 2016).

The study makes a significant theoretical and methodological contribution to the literature as the first paper to systematically test seasonality effects on preferences for beef cuts using a novel Discrete Choice Experiment (DCE). The study includes attributes related to the appearance such as meat and fat colour, fat content, packaging type as well as labelling information, including price, brand, origin, traceability, weight, type of feed, certification logos, expiry date, and claims regarding organic, angus, pasture raised, non-GMO, and natural beef. By systematically manipulating these variables through the DCE, the study recognizes the main factors that US consumer consider when purchasing beef products. The DCE is also innovative in replacing the typical "pick one" choice with quantities, i.e., how many units (including zero) of each product would you buy?

The significant empirical findings of this study also contribute to the management and practice of developing new products as well as improving the existing products. Furthermore, the findings will inform different functional departments within the food industry to effectively meet consumer needs (Ardeshiri & Rose, 2018; Jacobsen et al., 2014) The paper, therefore, concludes with a discussion of the implications of the findings for industry.

## 2. BEEF LABELLING AND CONSUMER PREFERENCES

In the US, labelling of meat products is closely regulated by the United States Department of Agriculture (USDA) and Food Safety and Inspection Service (FSIS). The FSIS has firm guidelines concerning the appearance and content of meat product labels. These strict labelling requirements, including country of origin labelling (COOL), attempts to provide the knowledge needed for consumers to make informed decisions (Ikenson, 2004; Jin, Skripnitchenko, & Koo, 2004; Umberger, 2004). The USDA has eight specific requirements for each product label which are: (1) product name, (2) inspection legend and establishment number, (3) handling statement, (4) net weight statement, (5) ingredients statement, (6) address line, (7) nutrition facts, and (8) safe handling instructions (US Department of Agriculture & Service, 2005). The restricted space available on the front of the package urge the producers to know exactly what additional information to the mandatory ones should be provided to maximise consumers preference for that product.

Consumers' value judgements related to product quality is based on intrinsic and extrinsic beef attributes (Asioli et al., 2017). Examples of intrinsic beef cues are, but not limited to, beef colour, cut of the meat, visible fat, and nutritional attributes (Acebrón & Dopico, 2000) and examples of extrinsic attributes are price, health related claims, brand, packaging layout (Jaeger, 2006; Lähteenmäki, 2013).

Extensive research has been conducted on the elicitation of US consumers' preferences for different beef products information cues. Intrinsic, quality traits such as the freshness, colour, and fat content of beef have been shown to influence US consumers' purchasing decisions. Numerous studies assessing aesthetic appeal and taste evaluations have shown that US, Italian and Spanish consumers prefer and believe red beef tastes superior to purple and brown cuts (Carpenter, Cornforth, & Whittier, 2001; Grebitus, Jensen, & Roosen, 2013; Realini et al., 2014; Zanoli et al., 2013). However, preferences differ between US consumers and consumers from other countries for other intrinsic attributes. Specifically, US consumers are more accepting of beef that has been fed genetically modified corn or has growth hormones added, compared to French, German and UK consumers (Lusk, Roosen, and Fox (2003). While US consumers may be willing to purchase corn-fed beef, evidence suggests they prefer grass-fed (Abidoye, Bulut, Lawrence, Mennecke, & Townsend, 2011; Lim & Hu, 2013).

Other studies have focused on the impact of extrinsic product information on consumer preferences, such as food safety, traceability and COOL. Studies have shown all three types of information have a significant impact on consumer preferences (Loureiro & Umberger, 2007). US consumers are willing to pay a premium for food safety information to be included on the package, such as the food borne disease risk (Lim & Hu, 2013), and whether the beef has been tested for bovine spongiform encephalopathy (BSE) (Lim & Hu, 2013; Lim, Hu, Maynard, & Goddard, 2012). Furthermore, the importance of food safety information is magnified for consumers with strong perceptions and attitudes towards risk (Lim et al., 2012). Traceability information and COOL also contributes to safety and quality perceptions, with various studies showing the presence of traceability information increases consumer preference for a beef product (Abidoye et al., 2011; Lim et al., 2012; Loureiro & Umberger, 2007).

US consumers are also willing to pay a premium for national/domestic produced beef products. Multiple studies have shown that US beef is preferred to beef imported from Canada or Mexico (Lim & Hu, 2013; Tonsor, Schroeder, & Lusk, 2013). This premium for US-labelled steak is as high as 20 percent in one study (Umberger, Feuz, Calkins, & Sitz (2003) and is likely due to poor food safety perceptions of imported beef Loureiro and Umberger (2003).

**3. DISCRETE CHOICE EXPERIMENT**

To understand consumer behaviour, DCE were used to analyse consumer choices. In these studies consumers behaviour is elucidated using either the Lancaster's (1966) consumer utility maximization model or McFadden (1974) random utility theory framework. Using the random utility theory, DCEs measure the relative importance of studied attributes by asking participants to repeatedly choose among given alternatives with different combination of attributes (J. Louviere & Hensher, 1982; J. Louviere, Hensher, & Swait, 2000). DCEs open up the possibility to explore the complex relationship between individual preferences and their values for different sets of attributes when making a decision and can be expanded across the populations of consumers (McEachern, Seaman, Padel, & Foster, 2005).

**3.1 EXPERIMENT DESIGN AND MATERIALS**

The study uses the novel choice modelling framework presented in Ardeshiri & Rose (2018) to uncover the effect of seasonality on consumer preferences for beef products. Participants were asked to select how many (including none) of each beef product would they most likely purchase. Nine different beef cuts were studied according the Institutional Meat Purchase Specifications (USDA, 2014) and North American Meat Processors Association (2007) descriptions. The attributes of ground, diced, roast, sirloin, tenderloin, flank, flap, New York strip and cowboy (rib eye)[1] were investigated in a DCE (refer to Figure 1 for a brief description of each cut). Feeding from the list of attributes presented in Ardeshiri & Rose (2018) study and conducting focus groups and interviews with experts in the US market and also looking at the literature review of similar studies conducted in the US, for each beef cut, this study used five discrete attributes (fat & meat Colour , marbling, type of packaging and brand) , three continues attributes (best before, weight and price attributes) and ten binary attributes (claim attributes) as presented in Table 1. Although the list of attributes selected is extensive and in addition to the restriction of having limited space available to present the information, one can argue that individuals may not asses all the information in their decision process, however, in the focus group, unanimously, it was agreed all the mentioned attributes is of importance in the decision process and should be investigated. A comparison among all attribute will also enable the producers to know exactly what additional information to the mandatory ones should be provided to maximise consumers preference for that product.

---

[1] The selection of the beef cuts was developed in consultation with the industry partner involved in the ARC grant associated with this research.

A key aspect in DCEs is the design of the set of choice alternatives that are presented to the respondents. DCEs consist of a sample of choice sets selected from the universal set of all possible choice sets that satisfy certain statistical properties. There is no consensus in the literature about the 'appropriate' number of choice sets per respondent. Louviere et al. (2000) noted that most studies ask respondents to evaluate between one and sixteen choice sets, with the average being somewhere around eight choice scenarios per respondent. In a later paper, (J. J. Louviere, 2004) argued that 'in contrast to the equivalent of widely held 'academic urban myth' in marketing and transport research, there is considerable evidence that humans will 'do' dozens (even hundreds) of T's [choice sets]' (p. 16). A number of studies have investigated the impact of the number of choice sets given to each respondent (Bech, Kjaer, & Lauridsen, 2011; Caussade, de Dios Ortúzar, Rizzi, & Hensher, 2005; David A Hensher, 2006; David A Hensher, Stopher, & Louviere, 2001; Hess, Hensher, Daly, & practice, 2012; Stopher & Hensher, 2000). Hess et al. (2012) argued that the concerns about fatigue in the literature are possibly overstated, with no clear decreasing trend in scale across choice tasks in any of their studies. They further stated that while the work by Bradley & Daly (1994) has become a standard reference in this context of reducing respondent engagement as a result of fatigue or boredom in large number of choice tasks, it should be recognised that not only was the fatigue part of the work based on a single dataset, but the state-of-the-art and the state-of-practice in stated choice survey design and implementation has moved on significantly since their study. Bech et al. (2011) studied the effect of 5,9 and 17 choice sets presented to each respondent and they found no differences in response rates and no systematic differences in the respondents' self-reported perception of the uncertainty of their DCE answers. Although there were some differences in willingness to pay (WTP) estimates suggesting that estimated preferences are to some extent context-dependent, but no differences in standard deviations for WTP estimates or goodness-of-fit statistics. Hensher et al. (2001)and Stopher and Hensher (2000) found that the number of choices had little impact on response rate, no impact on respondent fatigue and simplification of response strategies, minimal impact on the goodness of- fit statistics, and finally, little impact on the mean WTP estimates. Hensher (2006) concluded that the number of choice sets presented to the respondents did not influence the aggregate mean WTP estimates. Caussade et al. (2005) investigated five design parameters (number of alternatives in a choice set, number of attributes defined in each alternative, number of levels, range of attribute levels and number of choice sets) and concluded that all five parameters affected variance but did not systematically affect mean WTP estimates. Furthermore, as cited in Adamowicz, et al. (1998), a study conducted by Bunch and Batsell in 1989 demonstrated that with as few as six respondents per choice scenario, the asymptotic properties for maximum likelihood-based inference are satisfied (p. 15). For this study we used the Ngene software and applied an orthogonal main effects experimental design (J. Louviere et al., 2000), to select 200 choice task from the universal set for the experimental design. The 200 choice tasks were divided into 50 blocks allowing each participant to complete 4 repeated choice tasks. Four random alternatives were presented in each task that vary in the attribute levels. As shown in Table 1, the various attributes presented in all alternatives included; beef and fat colour, marbling (ground and diced beef were excluded), types of packaging, origin/brand[2], claims, weight and price. As illustrated in Figure 3, the traditional underlying mechanism of "pick a product" in traditional choice experiment surveys was substituted with "how many" (including zero) products would you purchase, to represent the real-life ordered logit structure when purchasing a food product (Ardeshiri & Rose, 2018). Each respondent was assigned to a winter or summer scenario and was asked to complete four sets. As shown in Figure 2, based on the season that was assigned to the individual, a current weather widget appeared next to the task to remind the individual about the hypothetical season they are shopping in.

## 3.2 DATA COLLECTION

Data for our analysis derived from an online survey distributed through a panel company in April 2017 and was completed by 946 American residents of north-eastern US. The studied regions included Connecticut, Maine, Massachusetts, New Hampshire, New Jersey, New York, Rhode Island and Vermont. From the 946 respondents, 468 were allocated a summer scenario and 478 were assigned to the winter scenario. Any US residents aged 18 years and above and responsible for their household grocery shopping, including the household's meat purchases were eligible to participate in the survey. The majority of respondents (39%) purchase beef once a week and only 16% purchase beef once a month or less. A summary of the sample demographics is provided in Table 2.

---

[2] For the design of this study country of origin and brands have been constrained and are coincident.

## 3.3 DATA ANALYSIS

It can reasonably hypothesize that each individual has a continues preference with different strength that underlines "how many" beef products they will purchase. This strength of preference is labelled as "utility" (*V)* and the utility value can have the following range of values:

$$-\infty < V_{im} \leq \infty$$

where *i* is referred to the individual and *m* presents the beef product. In this experiment the individuals had the option to select a value from 0 to 10 as an indication of "how many" of the product they would like to purchase. The underlying utility will then be translated to a rating and viewed as a *censoring* of the underlying utility,

$$
\begin{aligned}
y_{im} &= 0 \; if \; -\infty < V_{im} \leq \tau_{m1} \\
y_{im} &= 1 \; if \; \tau_{m1} < V_{im} \leq \tau_{m2} \\
&\vdots \\
y_{im} &= j \; if \; \tau_{mj} < V_{im} \leq \infty
\end{aligned}
\qquad (1)
$$

The thresholds, $\tau_{mj}$, are specific to the beef cut and number *(J-1)* where *J* is the number of possible ratings (here, eleven) *J-1* values are needed to divide the range of utility into *J* cells. (For a more detailed explanation of the threshold elements we encourage the readers to read Ardeshiri & Rose, 2018; Greene & Hensher, 2010).

The utility function includes individual preferences $\beta_i$ for beef attributes which we denote $x_{i1}, x_{i2}, \ldots, x_{iK}$. It also embraces individual sociodemographic variables $z_{i1}, z_{i2}, \ldots, z_{iK}$, to allow for covariate specific heterogeneity. And finally, an aggregate of unmeasured and unmeasurable (by the statistician) idiosyncrasies, denoted $\varepsilon_{im}$. For conventional reasons, we assume these features have a linear format in the utility function. The described utility function is presented as below:

$$U_{im} = ASC_i + \beta_i x_{ik} + \gamma_i z_{ik} + \varepsilon_{im} \quad (2)$$

Initially two separate ordered logits were estimated for summer and winter scenarios. In order to determine whether the preferences stated in both methods were proportionally similar, the coefficients derived from the two models were plotted (D. Hensher, Louviere, & Swait 1998; Huynh, Coast, Rose, Kinghorn, & Flynn, 2017; Swait & Bernardino, 2000). This offers a preliminary suggestion as to whether the data is likely to be possible to be pooled together. In other words, the respondents are using similar cognitive processes, although the decision context is changing. A Chow test for data pooling was conducted using the Swait and Louviere (1993) approach. This test compares the sum of the log likelihood from the ordered logit models of each season, to a log likelihood that allows for scale differences across the data sources (Swait, 2006). This will allow testing the hypothesis that there is homogeneity in preferences across the attributes included in the utility per seasons, whilst allowing for error variance differences between the unobserved factors in the two datasets. Passing this test at the 1% significance level indicates that it would be suitable to combine the datasets as long as the different scale factors are accounted for. The combined data were analysed using a seasonality scale[3] adjusted ordered logit model in Python Biogeme 2.4. (Bierlaire, 2016). The final utility function to be estimated for each decision maker *i* is:

(3)

---

[3] In this study eighteen scales (holding one constant at one) represented the nine studies cuts in two different seasons.

$$U_{im} = \exp(\mu_{c_b s_j}) [ASC_i + \beta_i x_{ik} + \gamma_i z_{ik} + \varepsilon_{im}]$$

Where $\mu_{c_b s_j}$ represents the scale value for cut $b$ and season $j$. Alternative specific beef attributes as well as season specific parameters were measured. However, some attributes entered the final model as generic across all or a range of beef products. The DCE attributes with qualitative levels are effects coded to ensure the systematic utility effects are uncorrelated with the intercept (Bech & Gyrd-Hansen, 2005; J. Louviere et al., 2000). Any coefficients that were not significant at the 90% level of accuracy have been removed using Louviere et al. (2000) log likelihood ratio test.

### 3.4 WILLINGNESS TO PAY ESTIMATION

Increasingly, discrete-choice models are being used to derive estimates of money an individual is willing to pay (or willing to accept) to obtain some benefit (or avoid some cost) from a specific action (J. Louviere et al., 2000). In the presented model where each attribute in a utility expression is associated with a single taste weight, the WTP for a unit change in a certain attribute can be computed as the marginal rate of substitution between cost and the quantity expressed by the attribute, holding all other potential influences constant.

$$WTP = -\frac{\beta_i}{\beta_{cost}} \tag{4}$$

WTP estimates for categorical attributes refer to the benefit obtained by having that level and for continues attributes, it represents the benefit of a change in a unit of the attribute.

## 4. RESULTS

Estimation results from the scaled adjusted ordered logit model are presented in several tables. Tables 3 and 4 present seasonal parameter estimates for specific beef cuts. Table 5 presents the covariate estimates and Table 6 presents threshold estimates.

Consumers showed similar taste preferences for beef colour, fat colour and marbling in both seasons. A generic parameter was used to measure beef and fat colour preferences across all beef cuts and both seasons. US consumers prefer white fat in comparison to light yellow, and red beef over pink. Consumers are willing to pay $3.14 and $2.18 respectively for beef cuts with white fat and red beef. An alternative specific parameter was used to represent marbling preferences across all beef cuts, however, the beef specific parameters are the same across both seasons (marbling does not apply to ground or diced beef). No differences were observed between no marbling and somewhat marbled levels for tenderloin, flank and New York steak. No marbling was preferred for roast, sirloin and flap, with a respective WTP of $10.18, $16.46 and $3.89. Somewhat marbled is more preferred for cowboy cuts and US consumers are willing to pay $17.92 for this.

An alternative specific parameter representing both beef cuts and season were used to present heterogeneity in preferences for packaging type. For all beef cuts, except cowboy, tray packed was less preferred relative to vacuumed packed and sold fresh. While this finding holds across both seasons, the magnitudes differ. For example, US consumers have a stronger negative preference for tray packed diced, tenderloin and flank beef in summer relative to winter. In general, consumers value fresh sold ground and diced beef; vacuum packed roast, sirloin, flank and New York steak; and tray packed cowboy steaks in both seasons. Vacuum packed tenderloin is preferred in winter and fresh in summer. Finally, no difference in packaging types were found for flap cuts in winter, although consumers prefer fresh in summer. For WTP estimates please refer to Tables 7 and 8. These results are in line and contribute (by observing preference for different beef cuts) to results from Carpenter et al. (2001) where they observed U.S. consumers preferred vacuum skin packaging.

With regards to the labelling information, the largest difference in preferences between seasons is for certified logos representing the authority which has approved and certified the beef. Consumers reacted strongly to packages with no certification logo in summer, compared to in winter. The highest negative WTP for no program or logo in winter was calculated as -$28.42 for New York strip, whereas in summer this value was calculated at -$87.05.

Consumer preferences for product claim information is homogenous for both seasons. Heterogeneity in preferences appears only in claims related to specific cuts of beef. Grass fed is preferred over grain fed and US consumers have the highest WTP for grass fed cowboy cuts at $11.60. Furthermore, US consumer are willing to pay a premium for beef products that are traceable back to farm, with the highest WTP for New York strip at $15.89. The impact of antibiotic free claims on WTP have the highest variance, ranging from $4.53 for ground beef up to $47.52 for sirloin steak. On the contrary, WTP for hormone free claims have the lowest variance, ranging from $0.69 for diced beef to $2.27 for sirloin steak. US consumers also prefer beef that is Angus, organic, non-GMO, pasture and natural (please view Tables 7 and 8 for all the WTP calculated values). Different to Grebitus, Jensen, & Roosen (2013), this study produced a lower WTP for longer use-by dates in summer than in winter.

US consumers are also willing to pay more for larger cuts of roast and flap in winter than in summer. Flank has a negative estimated coefficient for weight in both seasons, demonstrating that consumers prefer flank steaks in smaller portions. As anticipated, the price coefficient was negative for all beef products. US consumer are more sensitive to a price increase for diced products. Tables 7 and 8 provide all the WTP calculated values.

Table 5 presents the covariate coefficient estimates. The results from this table demonstrate that covariates such as having a graduate degree, owning a dwelling, larger household size, being a male, living in New York state, purchasing beef 2 plus times a week, belonging to a couple family with no kids' household and being in the lower age spectrum will increase the probability of purchasing beef products. Moreover, a quadratic form of income became significant with a negative value representing that middle-income US consumer will more likely purchase beef products rather than individuals belonging to both ends of the income range. The majority of the scale parameters are not statistically different from one at the 95% confidence level: only the scale estimates for ground, diced and roast in summer. This is to be expected as the scale differences have been captured through introducing season and beef specific parameters in the utility function.

Table 6 provides the estimation of the threshold properties. The thresholds are an important element of the model; they divide the range of utility into cells that are then identified with the observed ratings. One of the admittedly unrealistic assumptions in many applications is that these threshold values are the same for all individuals. Importantly, difference on a rating scale (e.g., 0 compared to 1, 1 compared to 2) are not equal on a utility scale; hence we have a strictly nonlinear transformation captured by the thresholds, which are estimable parameters in an ordered choice model (Ardeshiri & Rose, 2018; Greene & Hensher, 2010). The threshold parameters are all incrementally increasing, in order to preserve the positive signs of all of the probabilities. For example, in Table 6 the ground cut threshold (1) is equal to 0, threshold (2) is 1.57, threshold (3) is 2.74, etc.

## 5. DISCUSSION

Beef industries need to be ensured that their products match consumer demand and they retain their market share. The study outlined a choice experiment and attempt to differentiate between internally variety-seeking behaviour and seasonality as an external motivator.

The outcome of the study suggest that consumers vary in their preferences for beef products by season. The presence of a USDA certification logo is by far the most important factor affecting consumer's willingness to pay for all beef cuts, except flank. The value of USDA certification to consumers is heavily dependent on season, with customers willing to pay up to an additional $58 for USDA certification in summer for New York strip beef than in winter. Similarly, consumers are highly averse to purchasing beef with no certification (USDA or otherwise). This effect is stronger in summer, indicating that consumers are concerned about food safety and may be aware of the opportunities for foodborne bacteria to thrive in warm weather (Figure 4).

In relation to packaging, US consumers have mixed preference for different beef products by season. For ground and diced beef, consumers prefer to purchase the beef fresh over the counter in both seasons. Cowboy cut is the only product that consumers prefer to have in trey packed in both seasons. Although US consumers prefer New York strip beef in vacuum packed the most but they still value purchasing the product fresh over the counter in winter whereas in summer they only prefer in vacuumed packed.

The results also show that for some products, consumers differ in their preferences by season for product net weight. For example, the bigger the flap and roast products in winter the better. US consumers are willing to pay an extra $3.9 and $1.75 for flap and roast products in winter than in summer respectively. However, the opposite applies for Sirloin steak where US consumers are willing to pay an extra $2.9 for a higher product weight in summer than in winter. A higher product weight for flank is perceived negatively for both seasons and it intensifies by -$1.35 in summer relative to winter. Finally, having a longer used by date for Sirloin, NY strip and Flap is more preferable in winter than in summer. Moreover, for Flap cuts with having a longer used-by-date in summer is seen as a disadvantage.

Regardless of seasonality effect, the results of this study also support the findings of previous work in a number of ways. Claims are important to US consumers. Antibiotic-free is the most important claim for the roast ($10.59), sirloin ($47.50) and flank ($10.79), but it has no effect on willingness to pay for tenderloin beef ($0). Although no added hormones had a relatively small effect on willingness to pay, it had an effect on preferences across all cuts. Table 9 summarises the impact of each claim on the willingness to pay and for which cuts.

With regards to product appearance, white fat colour was preferred more than light yellow fat colour and consumers were WTP a premium of $1.61 for diced beef and up to $5.26 for sirloin. Similar to Zanoli et al. (2013) and Carpenter et al. (2001) study, red coloured beef was the most preferred beef colour. Sirloin steak with $3.66 has the highest premium for having a red coloured meat while diced beef with $1.12 has the lowest premium. Intramuscular fat content also known as marbling has been considered as one of the main determinants in the beef quality grading system. Parallel to the findings of Lusk et al. (2003), US consumers preferred beef products (not applicable to ground and diced beef products) with the least amount or no marbling. Sirloin steak with $16.46, received the highest premium for having no intramuscular whereas the lowest was for Cowboy cut with -$17.92.

Other non-DCE studies have indicated that consumers' preferences for purchasing beef products is also shaped by other variables such as the location; frequencies of eating beef (by specific cut) in restaurants, cafés, bars, etc.; cultural beliefs; level of knowledge; religious beliefs; environmental sustainability preferences and so on. Thus, a limitation of this study that should be addressed in future research is to explore and identify consumers segments to improve marketing strategies for beef producers based on other heterogeneity features available in segments of the population.

### 5.1 MANAGERIAL AND POLICY IMPLICATIONS

The current study benefits from inclusive range of policy implication influencing consumers, businesses and policy makers. Authorities and policy makers require information about consumer preferences to make informed decisions and avoid market failures. One of the reasons market failures occur is that the intrinsic and extrinsic attributes of a product do not meet consumers' needs, or, due to labelling, are not perceived by consumers to meet their needs. Consumers are faced with a large number of food choices on the daily bases and it is unlikely to allocate significant cognitive effort and time to each of their decisions. Therefore, it is important that products and labels reflect the information consumers value when purchasing beef. The findings of this study will help managers design labels with useful and meaningful information for consumers, as well as help consumers to navigate product choices more efficiently. Industry firms can profit from similar research with their product differentiation strategy, cost-benefit assessment and product design by season. The study also highlights the importance of seasonality effects on consumers purchasing behaviour.

#### 5.1.1. HYPOTHETICAL SCENARIO

This section provides an example of how the output of this research can benefit business firms with the decision making and market assessment on a newly developed product by season.

For this reason, Figures 5 and 6 presents purchase probabilities for all the nine cuts in both seasons. The product details (i.e. fat colour, beef colour, claims, brands and the used-by date) are constrained to be the same among all nine cuts. The average value for price per pound and package size differed based on the beef cut, however, they were assumed to be equal in both seasons. Figures 6 and 7 presents the probability of purchasing different quantities (including none) for winter and summer. Roast and diced have the biggest change in units purchased between seasons. In summer, roast and diced respectively have a 15% and 16% probability of having zero units purchased, compared to 10% and 11% in winter. This result suggests that winter is the slow cooked season. However, the simulation also shows that winter competes strongly with summer as the grilling season, as there are few differences between the summer and winter steak purchase probabilities: only New York strip has a higher probability of not being purchased in winter compared to summer (78% purchase probability of at least one unit in summer, compared to 74% in winter).

The average purchase quantity for each product supports the underlying comparison made earlier by only looking at the zero purchase probabilities. On average, diced beef has the highest purchase quantity in winter with more than 2 quantities and, as expected, tenderloin has the lowest purchase quantity in summer as it is the most expensive cut. Although flank steak had the same probability of zero purchases (10%), on average it is being purchased more in summer. This results in the New York strip and flank steaks to be the only grilling products to have a higher average purchase quantity in summer than in winter.

The above scenarios present a simple example of the benefit of this study to the beef industry. By creating hypothetical products, managers can observe and predict the attractiveness of a new product. Furthermore, by using the product weights, the market share of these new products can be estimated based on total pounds sold for each beef cut. The purchase probability trends can be drawn for each cut to observe the different probabilities for an incremental change in price values. These graphs can help to find the optimal price value that maximises revenue. This is a simple, yet powerful, predictive tool for beef producers, processors and policymakers to strengthen their decision-making capabilities.

## 6. CONCLUSION

This study provides information regarding the importance and effect of different information cues to US consumers when selecting beef products in winter and summer.

This study represents how choice experiments can offer valuable insights for businesses as well as policymakers to make inform decisions while considering how consumers relatively value among different labelling and product attributes by season and better address any ethical, safety and aesthetic concerns that consumers might have. Finally, the findings of this study can assist policymakers and stockholders with estimating the economic benefits of a given policy measure by season.

A limitation to this study and as a further research stream is to investigate consumers processing resources and cognitive efforts for each decision. Given the sheer number of decisions involved across the many facets of people's lives, it seems unlikely that individuals allocate substantial cognitive effort and time to each decision. Indeed, decisions regarding small budget items like food or consumer packaged goods would seem more likely to be relegated to some form of habitual choice behaviour (W. L. Adamowicz & Swait, 2012). Considering a large number of attributes included boredom and fatigue effects may be observed if respondents are presented large numbers of complex choice tasks. While literature supports the notion that increasing complexity over a number of attributes changes choice behaviour toward strategies that employ less attribute, this may not always be the case, of course (Swait & Adamowicz, 2001).

Finally, an alternative suggestion for future directions of research is to investigate and compare current market data and the observed market behaviour with the finding of this study to increase the robustness of the results.

# References:

Abidoye, B. O., Bulut, H., Lawrence, J. D., Mennecke, B., & Townsend, A. M. (2011). US consumers' valuation of quality attributes in beef products. *Journal of Agricultural and Applied Economics, 43*(1), 1.

Acebrón, L. B., & Dopico, D. C. (2000). The importance of intrinsic and extrinsic cues to expected and experienced quality: an empirical application for beef. *Food Quality and Preference, 11*(3), 229-238.

Adamowicz, W., Louviere, J., & Swait, J. J. N.-N. O. A. A., Washington, USA. (1998). Introduction to attribute-based stated choice methods.

Adamowicz, W. L., & Swait, J. D. (2012). Are food choices really habitual? Integrating habits, variety-seeking, and compensatory choice in a utility-maximizing framework. *American Journal of Agricultural Economics, 95*(1), 17-41.

Ardeshiri, A., & Rose, J. M. (2018). How Australian consumers value intrinsic and extrinsic attributes of beef products. *Food Quality and Preference, 65*, 146-163.

Asioli, D., Varela, P., Hersleth, M., Almli, V. L., Olsen, N. V., & Næs, T. (2017). A discussion of recent methodologies for combining sensory and extrinsic product properties in consumer studies. *Food Quality and Preference, 56*, 266-273.

Bech, M., & Gyrd-Hansen, D. (2005). Effects coding in discrete choice experiments. *Health economics, 14*(10), 1079-1083.

Bech, M., Kjaer, T., & Lauridsen, J. J. H. e. (2011). Does the number of choice sets matter? Results from a web survey applying a discrete choice experiment. *20*(3), 273-286.

Bierlaire, M. (2016). PythonBiogeme: a short introduction. Report TRANSP-OR 160706 ,Series on Biogeme. In: Transport and Mobility Laboratory, School of Architecture, Civil and Environmental Engineering, Ecole Polytechnique Fédérale de Lausanne, Switzerland.

Bradley, M., & Daly, A. (1994). Use of the logit scaling approach to test for rank-order and fatigue effects in stated preference data. *Transportation, 21*(2), 167-184.

Capps, O., Farris, D. E., Byrne, P. J., Namken, J. C., & Lambert, C. D. (1994). Determinants of wholesale beef-cut prices. *Journal of Agricultural and Applied Economics, 26*(1), 183-199.

Carpenter, C. E., Cornforth, D. P., & Whittier, D. (2001). Consumer preferences for beef color and packaging did not affect eating satisfaction. *Meat science, 57*(4), 359-363.

Caussade, S., de Dios Ortúzar, J., Rizzi, L. I., & Hensher, D. A. J. T. r. p. B. M. (2005). Assessing the influence of design dimensions on stated choice experiment estimates. *39*(7), 621-640.

Dijksterhuis, G. (2016). New product failure: Five potential sources discussed. *Trends in Food Science & Technology, 50*, 243-248.

Farris, D., & Holloway, D. (1990). *Demand Trends for Beef Cuts-by Quality, Convenience, and Season.* Paper presented at the American Journal of Agricultural Economics.

Grebitus, C., Jensen, H. H., & Roosen, J. (2013). US and German consumer preferences for ground beef packaged under a modified atmosphere–Different regulations, different behaviour? *Food policy, 40*, 109-118.

Greene, W. H., & Hensher, D. A. (2010). *Modeling ordered choices: A primer*: Cambridge University Press.

Hensher, D., Louviere, J., & Swait, J. (1998). Combining sources of preference data. *Journal of Econometrics, 89*(1), 197-221.

Hensher, D. A. (2006). Revealing differences in willingness to pay due to the dimensionality of stated choice designs: an initial assessment. *Environmental Resource Economics, 34*(1), 7-44.

Hensher, D. A., Stopher, P. R., & Louviere, J. J. J. J. o. A. T. M. (2001). An exploratory analysis of the effect of numbers of choice sets in designed choice experiments: an airline choice application. *7*(6), 373-379.

Hess, S., Hensher, D. A., Daly, A. J. T. r. p. A. p., & practice. (2012). Not bored yet–revisiting respondent fatigue in stated choice experiments. *46*(3), 626-644.

Hogan Jr, R. J., & Ward, C. E. (2003). *Models Estimating Beef Quality and Yield Grade Discounts.* Paper presented at the 2003 Annual Meeting, July 13-16, 2003, Denver, Colorado.

Huynh, E., Coast, J., Rose, J., Kinghorn, P., & Flynn, T. (2017). Values for the ICECAP-Supportive Care Measure (ICECAP-SCM) for use in economic evaluation at end of life. *Social Science & Medicine, 189*, 114-128.

Ikenson, D. (2004). Uncool Rules: Second Thoughts on Mandatory Country of Origin Labeling. *Free Trade Bulletin*(7).

Jacobsen, L. F., Grunert, K. G., Søndergaard, H. A., Steenbekkers, B., Dekker, M., & Lähteenmäki, L. (2014). Improving internal communication between marketing and technology functions for successful new food product development. *Trends in food science & technology, 37*(2), 106-114.

Jaeger, S. R. (2006). Non-sensory factors in sensory science research. *Food Quality and Preference, 17*(1), 132-144.

Jin, H. J., Skripnitchenko, A., & Koo, W. W. (2004). *The effects of the BSE outbreak in the United States on the beef and cattle industry* (Vol. 3): Center for Agricultural Policy and Trade Studies, Department of Agribusiness and Applied Economics, North Dakota State University Fargo.

Lähteenmäki, L. (2013). Claiming health in food products. *Food Quality and Preference, 27*(2), 196-201.

Lancaster, K. (1966). A new approach to consumer theory. *Journal of political Economy, 74*, 132-157.

Lim, K. H., & Hu, W. (2013). *How local is local? Consumer preference for steaks with different food mile implications.* Paper presented at the 2013 Annual Meeting, February 2-5, 2013, Orlando, Florida.

Lim, K. H., Hu, W., Maynard, L., & Goddard, E. (2012). *Stated Preference and Perception Analysis for Traceable and BSE-tested Beef: An Application of Mixed Error-Component Logit Model.* Paper presented at the 2012 Annual Meeting, August 12-14, 2012, Seattle, Washington.

Loureiro, M. L., & Umberger, W. J. (2003). Estimating consumer willingness to pay for country-of-origin labeling. *Journal of Agricultural and Resource Economics*, 287-301.

Loureiro, M. L., & Umberger, W. J. (2007). A choice experiment model for beef: What US consumer responses tell us about relative preferences for food safety, country-of-origin labeling and traceability. *Food policy, 32*(4), 496-514.

Louviere, J., & Hensher, D. A. (1982). On the Design and Analysis of Simulated Choice or Allocation Experiments in Travel Choice Modeling. *Transportation Research Record, 890*, 11-17.

Louviere, J., Hensher, D. A., & Swait, J. D. (2000). *Stated choice methods: analysis and applications*: Cambridge University Press.

Louviere, J. J. (2004). Random utility theory-based stated preference elicitation methods: applications in health economics with special reference to combining sources of preference data.

Lusk, J. L., Marsh, T. L., Schroeder, T. C., & Fox, J. A. (2001). Wholesale demand for USDA quality graded boxed beef and effects of seasonality. *Journal of Agricultural and Resource Economics*, 91-106.

Lusk, J. L., Roosen, J., & Fox, J. A. (2003). Demand for beef from cattle administered growth hormones or fed genetically modified corn: a comparison of consumers in France, Germany, the United Kingdom, and the United States. *American Journal of Agricultural Economics, 85*(1), 16-29.

McEachern, M., Seaman, C., Padel, S., & Foster, C. (2005). Exploring the gap between attitudes and behaviour: Understanding why consumers buy or do not buy organic food. *British food journal, 107*(8), 606-625.

Meat & Livestock Australia, M. (2017). *Australia - Beef export value - Calendar year*. Retrieved from: http://statistics.mla.com.au/Report/RunReport/45392dd9-4eae-426b-80e2-09993a85834c

Moskowitz, H., & Beckley, J. (2009). Seasonality and the algebra of food preferences revealed through product concepts. *Journal of sensory studies, 24*(1), 58-77.

Namken, J. C., Farris, D. E., & Capps Jr, O. (1994). The demand for wholesale beef cuts by season and trend. *Journal of Food Distribution Research, 25*(2), 47-61.

North American Meat Processors Association. (2007). *The meat buyer's guide: Beef, lamb, veal, pork, and poultry*.

Realini, C., Kallas, Z., Pérez-Juan, M., Gómez, I., Olleta, J., Beriain, M., . . . Sañudo, C. (2014). Relative importance of cues underlying Spanish consumers' beef choice and segmentation, and consumer liking of beef enriched with n-3 and CLA fatty acids. *Food Quality and Preference, 33*, 74-85.

Schroeder, T. C., & Yang, X. (2001). Hedging Wholesale Beef Cuts. *Journal of Agricultural and Resource Economics, 26*, 569-570.

Stopher, P., & Hensher, D. (2000). Are More Profiles Better Than Fewer?: Searching for Parsimony and Relevance in Stated Choice Experiments. *Transportation Research Board*(1719), 165-174.

Swait, J. (2006). Advanced choice models. In *Valuing environmental amenities using stated choice studies* (pp. 229-293): Springer.

Swait, J., & Adamowicz, W. (2001). The influence of task complexity on consumer choice: a latent class model of decision strategy switching. *Journal of Consumer Research, 28*(1), 135-148.

Swait, J., & Bernardino, A. (2000). Distinguishing taste variation from error structure in discrete choice data. *Transportation Research Part B: Methodological, 34*(1), 1-15.

Swait, J., & Louviere, J. (1993). The role of the scale parameter in the estimation and comparison of multinomial logit models. *Journal of Marketing Research*, 305-314.

Tonsor, G. T., Schroeder, T. C., & Lusk, J. L. (2013). Consumer valuation of alternative meat origin labels. *Journal of agricultural economics, 64*(3), 676-692.

Umberger, W. J. (2004). Will consumers pay a premium for country-of-origin labeled meat. *Choices, 19*(4), 15-19.

Umberger, W. J., Feuz, D. M., Calkins, C. R., & Sitz, B. M. (2003). Country-of-origin labeling of beef products: US consumers' perceptions. *Journal of Food Distribution Research, 34*(3), 103-116.

US Department of Agriculture, F. S., & Service, I. (2005). Food standards and labeling policy book. In: United States Department of Agriculture Washington, DC.

USDA. (2014). Institutional meat purchase specifications: Fresh beef series 100. In: USDA Washington, DC.

Zanoli, R., Scarpa, R., Napolitano, F., Piasentier, E., Naspetti, S., & Bruschi, V. (2013). Organic label as an identifier of environmentally related quality: A consumer choice experiment on beef in Italy. *Renewable Agriculture and Food Systems, 28*(01), 70-79.

**Table 1.** Attributes and levels in the choice experiment.

| Attribute | Levels | | |
|---|---|---|---|
| **Fat Colour*** | 1-White (0) 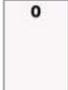 | 2-Light yellow (4) 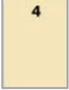 | |
| **Meat Colour*** | 1-Pink (1A) 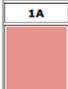 | 2-Red (3) 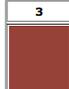 | |
| **Marbling*** | 1-Not marbled (0) 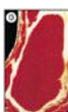 | 2-Somewhat marbled (4) 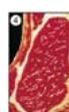 | |
| **Type of Packaging** | 1-Tray Packed (TP) Tray packed meat is when the meat is packed into an open container or tray and covered with a film. This is mainly used in smaller primal cuts or portioned meat. 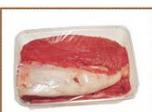 2-Vacuum Packed (VP) Vacuum Packed involves the removal of air and oxygen from the packaging. This creates a vacuum and assists in the preservation of meat and improvement in meat quality due to the lack of oxygen around the meat that promotes bacterial growth. 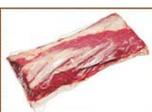 3- Fresh from the butcher. | | |
| **Brand**[4] | 1-Brand 1 4- Brand 4 7- Brand 7 | 2- Brand 2 5-Brand 5 8- Brand 8 | 3- Brand 3 6- Brand 6 |
| **Claim attributes** | 1-Grass fed 4- No added Antibiotic 7- No added Hormone 10- Natural | 2-Grain fed 5-Organic 8- Pasture Raised/ Not confined | 3-Traceable back to the farm 6-Angus 9- Non-GMO |
| **Best before** | 1- One day to expiry 4- Fourteen days to expiry | 2-Three days to expiry | 3-Seven days to expiry |
| **Weight** | Ground, diced & Flap: Roast: Sirloin: New York Strip & Flank: Tenderloin: Cowboy: | 12(OZ), 16(OZ) 3(Lb), 5(Lb) 8(OZ),12(OZ) 6(OZ),12(OZ) 4(OZ),9(OZ) 1(Lb), 1.5(Lb) | |
| **Price ($US/per lb)** | Ground, diced & Flank: Roast: Sirloin, Flap & Cowboy: New York Strip: Tenderloin: | $6, $12, $18, $24 $8, $14, $20, $26 $14, $20, $26, $32 $12, $18, $24, $30 $30, $42, $54, $66 | |

*Note that we used the handbook of Australian meat standards to present the levels for these attributes. The number between the parentheses refers to the reference standard score. Please see the following link for more information. https://www.ausmeat.com.au/custom-content/cdrom/Handbook-7th-edition/English/DA71F4DE-F68A-11DA-AA4B-000A95D14B6E.html

---

[4] Brands have been de-identified for confidentiality reason. Brands 1 & 7 are from Australia only, brands 2, 4 & 6 are from United States only and other brands are sourced from a mixture of countries.

**Table 2.** Demographic variables and summary statistics of choice experiment participants.

| Variable | Definition | Statistics |
|---|---|---|
| *Gender* | | |
| | Male | 50.4% |
| | Female | 49.6% |
| | Total participants | 946 |
| *Age* | | |
| | Modal age group | 25-34 years (28.1%) |
| *State of origin* | | |
| | New York | 72% |
| | Other | 28% |
| *Household type* | | |
| | Couple family with no children | 16% |
| | Couple family with children | 33.2% |
| | One parent family | 6.7% |
| | Single person household | 15.6% |
| *Household size* | | |
| | Average household size | 3.2 |
| *Household income* | | |
| | Modal income bracket | $75,000 to $100,000 (33.2%) |
| *Employment* | | |
| | Full-time | 58.8% |
| | Part-time | 15.1% |
| | Retired | 13.9% |
| | Un-employed | 12.2% |
| *Education* | | |
| | Graduate degree | 30% |
| | Bachelor's degree | 32% |
| | Associate's degree | 8% |
| | College graduate or less | 30% |
| *Dwelling status* | | |
| | Owned | 71.6% |
| | Renting | 28.4% |
| *Beef consumption frequency* | | |
| | 2+ per week | 24% |
| | Once a week | 39% |
| | 2 or 3 a month | 21% |
| | Once a month | 16% |

**Table 3** Ordered logit parameter estimate for *winter* season.

| Attribute | Level | Ground | Diced | Roast | Sirloin | Tenderloin | Flank | Flap | New York | Cowboy |
|---|---|---|---|---|---|---|---|---|---|---|
| **Scale** | (μ) | 1.18 | 1.32 | 1.413 | 1 | 1 | 1 | 1 | 1 | 1 |
| **ASC** | ASC | 0.278 | 0.237 | -0.193 | -0.178 | 0.341 | 0.957 | -0.772 | -0.806 | 0.136 |
| **Fat colour** | White | 0.022 | 0.022 | 0.022 | 0.022 | 0.022 | 0.022 | 0.022 | 0.022 | 0.022 |
| | Light Yellow* | -0.022 | -0.022 | -0.022 | -0.022 | -0.022 | -0.022 | -0.022 | -0.022 | -0.022 |
| **Meat colour** | Pink* | -0.015 | -0.015 | -0.015 | -0.015 | -0.015 | -0.015 | -0.015 | -0.015 | -0.015 |
| | Red | 0.015 | 0.015 | 0.015 | 0.015 | 0.015 | 0.015 | 0.015 | 0.015 | 0.015 |
| **Marbling** | Not marbled | - | - | 0.065 | 0.068 | - | - | 0.041 | - | -0.140 |
| | Somewhat marbled* | - | - | -0.065 | -0.068 | - | - | -0.041 | - | 0.140 |
| **Packaging type** | Vacuum Packed | - | 0.033 | 0.033 | 0.137 | 0.077 | 0.233 | - | 0.061 | -0.037 |
| | Tray packed | -0.024 | -0.072 | -0.041 | -0.125 | -0.140 | -0.418 | - | -0.072 | 0.090 |
| | Fresh* | 0.024 | 0.039 | 0.008 | -0.012 | 0.063 | 0.185 | - | 0.011 | -0.054 |
| **Feed type** | Grass | - | 0.043 | 0.038 | - | 0.034 | 0.033 | 0.048 | - | 0.091 |
| | Grain* | - | -0.043 | -0.038 | - | -0.034 | -0.033 | -0.048 | - | -0.091 |
| **Traceable back to farm** | Yes | 0.041 | 0.041 | 0.055 | 0.032 | - | 0.039 | 0.065 | 0.070 | 0.055 |
| | No* | -0.041 | -0.041 | -0.055 | -0.032 | - | -0.039 | -0.065 | -0.070 | -0.055 |
| **Antibiotic free** | Yes | 0.031 | 0.068 | 0.068 | 0.196 | - | 0.075 | 0.052 | - | 0.045 |
| | No* | -0.031 | -0.068 | -0.068 | -0.196 | - | -0.075 | -0.052 | - | -0.045 |
| **Hormone added** | Yes* | -0.009 | -0.009 | -0.009 | -0.009 | -0.009 | -0.009 | -0.009 | -0.009 | -0.009 |
| | No | 0.009 | 0.009 | 0.009 | 0.009 | 0.009 | 0.009 | 0.009 | 0.009 | 0.009 |
| **Organic** | Yes* | - | 0.055 | 0.064 | 0.057 | - | 0.067 | 0.026 | 0.075 | 0.065 |
| | No | - | -0.055 | -0.064 | -0.057 | - | -0.067 | -0.026 | -0.075 | -0.065 |
| **Angus** | Yes | 0.034 | 0.068 | 0.062 | 0.036 | 0.147 | - | - | 0.055 | 0.048 |
| | No* | -0.034 | -0.068 | -0.062 | -0.036 | -0.147 | - | - | -0.055 | -0.048 |
| **Non-GMO** | Yes* | 0.029 | 0.029 | 0.029 | 0.029 | 0.029 | 0.029 | 0.029 | 0.029 | 0.029 |
| | No | -0.029 | -0.029 | -0.029 | -0.029 | -0.029 | -0.029 | -0.029 | -0.029 | -0.029 |
| **Pasture** | Yes | - | - | - | 0.070 | 0.088 | 0.058 | 0.058 | 0.087 | 0.039 |
| | No* | - | - | - | -0.070 | -0.088 | -0.058 | -0.058 | -0.087 | -0.039 |
| **Natural** | Yes | - | - | - | - | 0.054 | - | - | 0.047 | 0.063 |
| | No* | - | - | - | - | -0.054 | - | - | -0.047 | -0.063 |
| **Certified logo** | USDA Verified* | 0.133 | 0.198 | 0.086 | 0.096 | 0.127 | 0.158 | 0.067 | 0.074 | 0.144 |
| | Global Animal Partnership | -0.04 | - | - | - | - | - | - | - | - |
| | Self-Assurance Program | - | - | - | - | - | -0.068 | - | 0.051 | -0.045 |
| | No Program/Logo | -0.094 | -0.198 | -0.086 | -0.096 | -0.127 | -0.09 | -0.067 | -0.125 | -0.099 |
| **Brands** | Brand 1* | 0.053 | 0.128 | 0.013 | -0.062 | -0.113 | -0.039 | -0.007 | 0.046 | 0.055 |
| | Brand 2 | -0.053 | -0.155 | - | - | -0.062 | - | -0.077 | -0.046 | 0.057 |
| | Brand 3 | - | -0.069 | -0.099 | - | 0.065 | - | - | - | - |
| | Brand 4 | - | 0.124 | - | 0.062 | 0.110 | - | 0.084 | - | - |
| | Brand 5 | - | - | - | - | - | - | - | - | -0.111 |
| | Brand 6 | - | 0.052 | - | - | - | - | - | - | - |
| | Brand 7 | - | -0.080 | - | - | - | 0.039 | - | - | - |
| | Brand 8 | - | - | 0.086 | - | - | - | - | - | - |
| **Use-By date** | Continues | 0.013 | 0.016 | 0.018 | 0.024 | 0.010 | 0.017 | 0.072 | 0.034 | 0.016 |
| **Net weight** | Continues | - | - | 0.065 | 0.021 | 0.018 | -0.082 | 0.064 | 0.010 | - |
| **Price** | Continues | -0.007 | -0.014 | -0.006 | -0.004 | -0.009 | -0.007 | -0.011 | -0.004 | -0.008 |

*Represents the base level

**Table 4** Ordered logit parameter estimates for *summer* season

| Attribute | Level | Ground | Diced | Roast | Sirloin | Tenderloin | Flank | Flap | New York | Cowboy |
|---|---|---|---|---|---|---|---|---|---|---|
| **Scale** | (μ) | 1 | 1 | 1 | 1 | 1 | 1 | 1 | 1 | 1* |
| **ASC** | ASC | 0.175 | 0.132 | -0.284 | -0.39 | 0.0713 | 0.978 | -0.0582 | -0.6340 | 0.0099 |
| **Fat colour** | White | 0.022 | 0.022 | 0.022 | 0.022 | 0.022 | 0.022 | 0.022 | 0.022 | 0.022 |
| | Light Yellow* | -0.022 | -0.022 | -0.022 | -0.022 | -0.022 | -0.022 | -0.022 | -0.022 | -0.022 |
| **Meat colour** | Pink* | -0.015 | -0.015 | -0.015 | -0.015 | -0.015 | -0.015 | -0.015 | -0.015 | -0.015 |
| | Red | 0.015 | 0.015 | 0.015 | 0.015 | 0.015 | 0.015 | 0.015 | 0.015 | 0.015 |
| **Marbling** | Not marbled | - | - | 0.065 | 0.068 | - | - | 0.041 | - | -0.140 |
| | Somewhat marbled* | - | - | -0.065 | -0.068 | - | - | -0.041 | - | 0.140 |
| **Packaging type** | Vacuum Packed | - | 0.033 | 0.033 | 0.114 | 0.077 | 0.242 | - | 0.043 | -0.037 |
| | Tray packed | -0.024 | -0.111 | -0.041 | -0.096 | -0.201 | -0.457 | -0.056 | - | 0.090 |
| | Fresh* | 0.024 | 0.078 | 0.008 | -0.018 | 0.124 | 0.215 | 0.056 | -0.043 | -0.054 |
| **Feed type** | Grass | - | 0.043 | 0.038 | - | 0.034 | 0.033 | 0.048 | - | 0.091 |
| | Grain* | - | -0.043 | -0.038 | - | -0.034 | -0.033 | -0.048 | - | -0.091 |
| **Traceable back to farm** | Yes | 0.041 | 0.041 | 0.055 | 0.032 | - | 0.039 | 0.065 | 0.070 | 0.055 |
| | No* | -0.041 | -0.041 | -0.055 | -0.032 | - | -0.039 | -0.065 | -0.070 | -0.055 |
| **Antibiotic free** | Yes | 0.031 | 0.068 | 0.068 | 0.196 | - | 0.075 | 0.052 | - | 0.045 |
| | No* | -0.031 | -0.068 | -0.068 | -0.196 | - | -0.075 | -0.052 | - | -0.045 |
| **Hormone added** | Yes* | -0.009 | -0.009 | -0.009 | -0.009 | -0.009 | -0.009 | -0.009 | -0.009 | -0.009 |
| | No | 0.009 | 0.009 | 0.009 | 0.009 | 0.009 | 0.009 | 0.009 | 0.009 | 0.009 |
| **Organic** | Yes* | - | 0.055 | 0.064 | 0.057 | - | 0.067 | 0.026 | 0.075 | 0.065 |
| | No | - | -0.055 | -0.064 | -0.057 | - | -0.067 | -0.026 | -0.075 | -0.065 |
| **Angus** | Yes | 0.034 | 0.068 | 0.062 | 0.036 | 0.147 | - | - | 0.055 | 0.048 |
| | No* | -0.034 | -0.068 | -0.062 | -0.036 | -0.147 | - | - | -0.055 | -0.048 |
| **Non-GMO** | Yes* | 0.029 | 0.029 | 0.029 | 0.029 | 0.029 | 0.029 | 0.029 | 0.029 | 0.029 |
| | No | -0.029 | -0.029 | -0.029 | -0.029 | -0.029 | -0.029 | -0.029 | -0.029 | -0.029 |
| **Pasture** | Yes | - | - | - | 0.070 | 0.088 | 0.058 | 0.058 | 0.087 | 0.039 |
| | No* | - | - | - | -0.070 | -0.088 | -0.058 | -0.058 | -0.087 | -0.039 |
| **Natural** | Yes | - | - | - | - | 0.054 | - | - | 0.047 | 0.063 |
| | No* | - | - | - | - | -0.054 | - | - | -0.047 | -0.063 |
| **Certified logo** | USDA Verified* | 0.282 | 0.27 | 0.223 | 0.21 | 0.254 | 0.258 | 0.265 | 0.332 | 0.255 |
| | Global Animal Partnership | -0.04 | - | - | - | - | - | - | - | - |
| | Self-Assurance Program | - | - | - | - | - | -0.068 | - | 0.051 | -0.045 |
| | No Program/Logo | -0.242 | -0.27 | -0.223 | -0.21 | -0.254 | -0.19 | -0.265 | -0.383 | -0.21 |
| **Brands** | Brand 1* | 0.053 | 0.075 | 0.013 | -0.062 | -0.048 | -0.039 | 0.048 | 0.046 | 0.055 |
| | Brand 2 | -0.053 | -0.127 | - | - | -0.062 | - | -0.132 | -0.046 | 0.057 |
| | Brand 3 | - | - | -0.099 | - | - | - | - | - | - |
| | Brand 4 | - | - | - | 0.062 | 0.110 | - | 0.084 | - | - |
| | Brand 5 | - | - | - | - | - | - | - | - | -0.111 |
| | Brand 6 | - | 0.052 | - | - | - | - | - | - | - |
| | Brand 7 | - | - | - | - | - | 0.039 | - | - | - |
| | Brand 8 | - | - | 0.086 | - | - | - | - | - | - |
| **Use-By date** | Continues | 0.013 | 0.016 | 0.018 | 0.011 | 0.010 | 0.017 | -0.018 | 0.009 | 0.016 |
| **Net weight** | Continues | - | - | 0.054 | 0.033 | 0.018 | -0.091 | 0.023 | 0.010 | - |
| **Price** | Continues | -0.007 | -0.014 | -0.006 | -0.004 | -0.009 | -0.007 | -0.011 | -0.004 | -0.008 |

*Represents the base level

**Table 5** Covariate estimates in the ordered logit model.

| Demographics | Description | Values |
|---|---|---|
| **Education** | Graduate degree | 0.023 |
| | Bachelor's degree | -0.047 |
| | Associate's degree* | 0.024 |
| **Dwelling** | Owned | 0.033 |
| | Renting* | -0.033 |
| **Household Size** | Continues | 0.014 |
| **Income** | Continues (Quadratic form) | -0.008 |
| **Origin** | New York state | 0.035 |
| | Other states* | -0.035 |
| **Frequency of purchase** | 2+ per week | 0.059 |
| | Once a month* | -0.059 |
| **Gender** | Female | -0.021 |
| | Male* | 0.021 |
| **Age** | Continues | -0.034 |
| **Household type** | Couple family with no children | 0.099 |
| | One parent family | -0.040 |
| | Couple family with children* | -0.059 |

*Represents the base level

**Table 6** Estimation of threshold properties in the ordered logit model.

| Threshold's (τ) | Ground | Diced | Roast | Sirloin | Tenderloin | Flank | Flap | New York | Cowboy |
|---|---|---|---|---|---|---|---|---|---|
| Threshold 1 | 0 | 0 | 0 | 0 | 0 | 0 | 0 | 0 | 0 |
| Threshold 2 | 1.57 | 1.74 | 1.61 | 1.9 | 3.04 | 1.83 | 1.87 | 1.8 | 1.63 |
| Threshold 3 | 2.74 | 2.96 | 3.09 | 3.31 | 4.1 | 3.54 | 3.11 | 3.18 | 2.95 |
| Threshold 4 | 3.61 | 3.88 | 4.01 | 4.21 | 4.94 | 4.84 | 4.13 | 3.94 | 3.81 |
| Threshold 5 | 4.56 | 4.71 | 4.92 | 5.03 | 5.85 | 5.87 | 5 | 4.68 | 4.59 |
| Threshold 6 | 5.28 | 5.49 | 5.99 | 5.77 | 6.24 | 6.79 | 5.48 | 5.39 | 5.42 |
| Threshold 7 | 5.78 | 6.22 | 6.32 | 6.38 | 6.62 | 7.36 | 6 | 5.73 | 5.85 |
| Threshold 8 | 6.35 | 6.98 | 6.8 | 6.99 | 7.07 | 7.91 | 6.54 | 6.49 | 6.58 |
| Threshold 9 | 6.76 | 7.39 | 7.21 | 7.55 | 7.63 | 8.25 | 7.01 | 7.19 | 6.99 |
| Threshold 10 | 7.46 | 8.09 | 7.9 | 8.25 | 8.33 | 9.17 | 7.77 | 8.11 | 7.68 |

**Estimation Report**

*Final log likelihood*    -18789.548

*Number of parameters*    257

Sample size    3784

**Table 7** Willingness to pay estimates for *winter* season.

| Attribute | Level | Ground | Diced | Roast | Sirloin | Tenderloin | Flank | Flap | New York | Cowboy |
|---|---|---|---|---|---|---|---|---|---|---|
| **Fat colour** | *White* | $3.14 | $1.61 | $3.40 | $5.26 | $2.48 | $3.14 | $2.05 | $4.93 | $2.78 |
| | *Light Yellow\** | -$3.14 | -$1.61 | -$3.40 | -$5.26 | -$2.48 | -$3.14 | -$2.05 | -$4.93 | -$2.78 |
| **Meat colour** | *Pink\** | -$2.18 | -$1.12 | -$2.37 | -$3.66 | -$1.73 | -$2.18 | -$1.43 | -$3.43 | -$1.93 |
| | *Red* | $2.18 | $1.12 | $2.37 | $3.66 | $1.73 | $2.18 | $1.43 | $3.43 | $1.93 |
| **Marbling** | *Not marbled* | - | - | $10.18 | $16.46 | - | - | $3.89 | - | -$17.92 |
| | *Somewhat marbled\** | - | - | -$10.18 | -$16.46 | - | - | -$3.89 | - | $17.92 |
| **Packaging type** | *Vacuum Packed* | - | $2.44 | $5.11 | $33.21 | $8.78 | $33.69 | - | $13.82 | -$4.68 |
| | *Tray packed* | -$3.47 | -$5.32 | -$6.36 | -$30.30 | -$16.01 | -$60.43 | - | -$16.41 | $11.57 |
| | *Fresh\** | $3.47 | $2.88 | $1.25 | -$2.91 | $7.23 | $26.75 | - | $2.59 | -$6.89 |
| **Feed type** | *Grass* | - | $3.17 | $5.89 | - | $3.93 | $4.71 | $4.52 | - | $11.60 |
| | *Grain\** | - | -$3.17 | -$5.89 | - | -$3.93 | -$4.71 | -$4.52 | - | -$11.60 |
| **Traceable back to farm** | *Yes* | $5.91 | $3.00 | $8.68 | $7.81 | - | $5.62 | $6.14 | $15.89 | $6.98 |
| | *No\** | -$5.91 | -$3.00 | -$8.68 | -$7.81 | - | -$5.62 | -$6.14 | -$15.89 | -$6.98 |
| **Antibiotic free** | *Yes* | $4.53 | $5.07 | $10.59 | $47.52 | - | $10.79 | $4.87 | - | $5.80 |
| | *No\** | -$4.53 | -$5.07 | -$10.59 | -$47.52 | - | -$10.79 | -$4.87 | - | -$5.80 |
| **Hormone added** | *Yes\** | -$1.35 | -$0.69 | -$1.47 | -$2.27 | -$1.07 | -$1.35 | -$0.88 | -$2.13 | -$1.20 |
| | *No* | $1.35 | $0.69 | $1.47 | $2.27 | $1.07 | $1.35 | $0.88 | $2.13 | $1.20 |
| **Organic** | *Yes\** | - | $4.07 | $10.09 | $13.92 | - | $9.74 | $2.47 | $17.02 | $8.33 |
| | *No* | - | -$4.07 | -$10.09 | -$13.92 | - | -$9.74 | -$2.47 | -$17.02 | -$8.33 |
| **Angus** | *Yes* | $4.87 | $5.04 | $9.65 | $8.68 | $16.81 | - | - | $12.45 | $6.16 |
| | *No\** | -$4.87 | -$5.04 | -$9.65 | -$8.68 | -$16.81 | - | - | -$12.45 | -$6.16 |
| **Non-GMO** | *Yes\** | $4.19 | $2.15 | $4.54 | $7.03 | $3.32 | $4.19 | $2.74 | $6.59 | $3.71 |
| | *No* | -$4.19 | -$2.15 | -$4.54 | -$7.03 | -$3.32 | -$4.19 | -$2.74 | -$6.59 | -$3.71 |
| **Pasture** | *Yes* | - | - | - | $16.90 | $10.05 | $8.34 | $5.43 | $19.66 | $4.94 |
| | *No\** | - | - | - | -$16.90 | -$10.05 | -$8.34 | -$5.43 | -$19.66 | -$4.94 |
| **Natural** | *Yes* | - | - | - | - | $6.21 | - | - | $10.70 | $8.06 |
| | *No\** | - | - | - | - | -$6.21 | - | - | -$10.70 | -$8.06 |
| **Certified logo** | *USDA Verified\** | $19.23 | $14.67 | $13.47 | $23.27 | $14.53 | $22.84 | $6.32 | $16.82 | $18.43 |
| | *Global Animal Partnership* | -$5.78 | - | - | - | - | - | - | - | - |
| | *Self-Assurance Program* | - | - | - | - | - | -$9.83 | - | $11.59 | -$5.76 |
| | *No Program/Logo* | -$13.59 | -$14.67 | -$13.47 | -$23.27 | -$14.53 | -$13.01 | -$6.32 | -$28.41 | -$12.67 |
| **Brands** | *Brand 1\** | $7.62 | $9.46 | $1.97 | -$15.10 | -$12.94 | -$5.65 | -$0.67 | $10.55 | $6.98 |
| | *Brand 2* | -$7.62 | -$11.48 | - | - | -$7.10 | - | -$7.22 | -$10.55 | $7.23 |
| | *Brand 3* | - | -$5.13 | -$15.51 | - | $7.46 | - | - | - | - |
| | *Brand 4* | - | $9.19 | - | $15.10 | $12.58 | - | $7.89 | - | - |
| | *Brand 5* | - | - | - | - | - | - | - | - | -$14.21 |
| | *Brand 6* | - | $3.86 | - | - | - | - | - | - | - |
| | *Brand 7* | - | -$5.90 | - | - | - | $5.65 | - | - | - |
| | *Brand 8* | - | - | $13.53 | - | - | - | - | - | - |
| **Use-By date** | *Continues* | $1.81 | $1.15 | $2.85 | $5.70 | $1.17 | $2.50 | $6.82 | $7.82 | $2.05 |
| **Net weight** | *Continues* | - | - | $10.15 | $5.05 | $2.02 | -$11.84 | $6.08 | $2.27 | - |

**Table 8** Willingness to pay estimates for *summer* season.

| Attribute | Level | Ground | Diced | Roast | Sirloin | Tenderloin | Flank | Flap | New York | Cowboy |
|---|---|---|---|---|---|---|---|---|---|---|
| **Fat colour** | *White* | $3.14 | $1.61 | $3.40 | $5.26 | $2.48 | $3.14 | $2.05 | $4.93 | $2.78 |
| | *Light Yellow\** | -$3.14 | -$1.61 | -$3.40 | -$5.26 | -$2.48 | -$3.14 | -$2.05 | -$4.93 | -$2.78 |
| **Meat colour** | *Pink\** | -$2.18 | -$1.12 | -$2.37 | -$3.66 | -$1.73 | -$2.18 | -$1.43 | -$3.43 | -$1.93 |
| | *Red* | $2.18 | $1.12 | $2.37 | $3.66 | $1.73 | $2.18 | $1.43 | $3.43 | $1.93 |
| **Marbling** | *Not marbled* | - | - | $10.18 | $16.46 | - | - | $3.89 | - | -$17.92 |
| | *Somewhat marbled\** | - | - | -$10.18 | -$16.46 | - | - | -$3.89 | - | $17.92 |
| **Packaging type** | *Vacuum Packed* | - | $2.44 | $5.11 | $27.64 | $8.78 | $34.99 | - | $9.75 | -$4.68 |
| | *Tray packed* | -$3.47 | -$8.22 | -$6.36 | -$23.20 | -$22.99 | -$66.07 | -$5.28 | - | $11.57 |
| | *Fresh\** | $3.47 | $5.79 | $1.25 | -$4.44 | $14.21 | $31.08 | $5.28 | -$9.75 | -$6.89 |
| **Feed type** | *Grass* | - | $3.17 | $5.89 | - | $3.93 | $4.71 | $4.52 | - | $11.60 |
| | *Grain\** | - | -$3.17 | -$5.89 | - | -$3.93 | -$4.71 | -$4.52 | - | -$11.60 |
| **Traceable back to farm** | *Yes* | $5.91 | $3.00 | $8.68 | $7.81 | - | $5.62 | $6.14 | $15.89 | $6.98 |
| | *No\** | -$5.91 | -$3.00 | -$8.68 | -$7.81 | - | -$5.62 | -$6.14 | -$15.89 | -$6.98 |
| **Antibiotic free** | *Yes* | $4.53 | $5.07 | $10.59 | $47.52 | - | $10.79 | $4.87 | - | $5.80 |
| | *No\** | -$4.53 | -$5.07 | -$10.59 | -$47.52 | - | -$10.79 | -$4.87 | - | -$5.80 |
| **Hormone added** | *Yes\** | -$1.35 | -$0.69 | -$1.47 | -$2.27 | -$1.07 | -$1.35 | -$0.88 | -$2.13 | -$1.20 |
| | *No* | $1.35 | $0.69 | $1.47 | $2.27 | $1.07 | $1.35 | $0.88 | $2.13 | $1.20 |
| **Organic** | *Yes\** | - | $4.07 | $10.09 | $13.92 | - | $9.74 | $2.47 | $17.02 | $8.33 |
| | *No* | - | -$4.07 | -$10.09 | -$13.92 | - | -$9.74 | -$2.47 | -$17.02 | -$8.33 |
| **Angus** | *Yes* | $4.87 | $5.04 | $9.65 | $8.68 | $16.81 | - | - | $12.45 | $6.16 |
| | *No\** | -$4.87 | -$5.04 | -$9.65 | -$8.68 | -$16.81 | - | - | -$12.45 | -$6.16 |
| **Non-GMO** | *Yes\** | $4.19 | $2.15 | $4.54 | $7.03 | $3.32 | $4.19 | $2.74 | $6.59 | $3.71 |
| | *No* | -$4.19 | -$2.15 | -$4.54 | -$7.03 | -$3.32 | -$4.19 | -$2.74 | -$6.59 | -$3.71 |
| **Pasture** | *Yes* | - | - | - | $16.90 | $10.05 | $8.34 | $5.43 | $19.66 | $4.94 |
| | *No\** | - | - | - | -$16.90 | -$10.05 | -$8.34 | -$5.43 | -$19.66 | -$4.94 |
| **Natural** | *Yes* | - | - | - | - | $6.21 | - | - | $10.70 | $8.06 |
| | *No\** | - | - | - | - | -$6.21 | - | - | -$10.70 | -$8.06 |
| **Certified logo** | *USDA Verified\** | $40.77 | $20.00 | $34.93 | $50.91 | $29.05 | $37.30 | $25.01 | $75.45 | $32.64 |
| | *Global Animal Partnership* | -$5.78 | - | - | - | - | - | - | - | - |
| | *Self-Assurance Program* | - | - | - | - | - | -$9.83 | $0.00 | $11.59 | -$5.76 |
| | *No Program/Logo* | -$34.99 | -$20.00 | -$34.93 | -$50.91 | -$29.05 | -$27.47 | -$25.01 | -$87.05 | -$26.88 |
| **Brands** | *Brand 1\** | $7.62 | $5.55 | $1.97 | -$15.10 | -$5.48 | -$5.65 | $4.57 | $10.55 | $6.98 |
| | *Brand 2* | -$7.62 | -$9.41 | - | - | -$7.10 | - | -$12.46 | -$10.55 | $7.23 |
| | *Brand 3* | - | - | -$15.51 | - | - | - | - | - | - |
| | *Brand 4* | - | - | - | $15.10 | $12.58 | - | $7.89 | - | - |
| | *Brand 5* | - | - | - | - | - | - | - | - | -$14.21 |
| | *Brand 6* | - | $3.86 | - | - | - | - | - | - | - |
| | *Brand 7* | - | - | - | - | - | $5.65 | - | - | - |
| | *Brand 8* | - | - | $13.53 | - | - | - | - | - | - |
| **Use-By date** | *Continues* | $1.81 | $1.15 | $2.85 | $2.74 | $1.17 | $2.50 | -$1.67 | $2.11 | $2.05 |
| **Net weight** | *Continues* | - | - | $8.41 | $7.95 | $2.02 | -$13.19 | $2.19 | $2.27 | - |

**Table 9:** Willingness to pay range between cuts for each product claim.

| Claims | Willingness to Pay range | Cuts* |
|---|---|---|
| Antibiotic free | $0.00-$47.50 | Cowboy, Diced, Flank, Flap, Ground, Roast, Sirloin |
| Pasture raised/not confined | $0.00-$19.66 | Cowboy, Flank, Flap, Sirloin, Tenderloin |
| Organic | $0.00-$17.00 | Cowboy, Diced, Flank, Flap, NY Strip, Roast, Sirloin |
| Angus | $0.00-$16.80 | Cowboy, Diced, Ground, NY Strip, Roast, Sirloin, Tenderloin |
| Grass Fed | $0.00-$11.60 | Cowboy, Diced, Flank, Flap, Roast, Tenderloin |
| Natural | $0.00-$10.70 | Cowboy, NY Strip, Roast, Tenderloin |
| Non-GMO | $0.00-$7.05 | Cowboy, Diced, Flank, Flap, Ground, NY Strip, Sirloin, Tenderloin |
| Traceable back to farm | $0.00- $5.90 | Cowboy, Diced, Flank, Flap, Ground, NY Strip, Roast, Sirloin, |
| Hormone Free | $0.70-$2.30 | Cowboy, Diced, Flank, Flap, Ground, NY Strip, Roast, Sirloin, Tenderloin |

* Cuts refers to the cuts with a willingness to pay above $0.00 for the presence of this claim.

| | |
|---|---|
| Ground Beef (IMPS 136), unless otherwise specified, ground beef may be derived from boneless meat which has been frozen and stockpiled. | 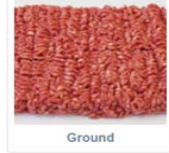 |
| Diced beef (IMPS/135) shall be prepared from any portion of the carcass which yields product that meets the end item requirements. | 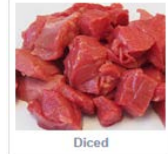 |
| Roast beef (IMPS/109) is that portion of the forequarter remaining after removal of the cross-cut chuck and short plate. | 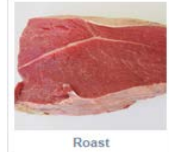 |
| Sirloin steak (IMPS/181) is the posterior section of the full loin. | 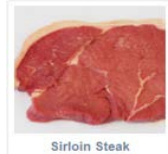 |
| NY strip steak (IMPS/158B) consists of the round (top, bottom, heel, rump, and shank) excluding the full sirloin tip (knuckle). | 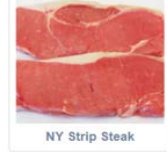 |
| Cowboy Steak (IMPS/1103B) may be prepared from any IMPS bone-in rib item. Each steak must be cut between the rib's bones. | 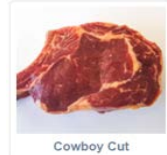 |
| Beef Tenderloin (IMPS/191) consist of the sirloin butt portion of the tenderloin. | 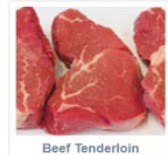 |
| Flank (IMPS/193) is a single flat muscle cut from the flank region of the carcass. | 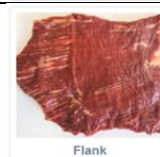 |
| Flap steak (IMPS/185A) comes from a bottom sirloin butt cut of beef and is generally a very thin steak. | 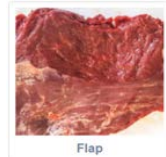 |

*Source:* Institutional Meat Purchase Specifications, 2014
https://www.ams.usda.gov/sites/default/files/media/IMPS_100_Fresh_Beef%5B1%5D.pdf

**Figure 1:** Beef cuts descriptions

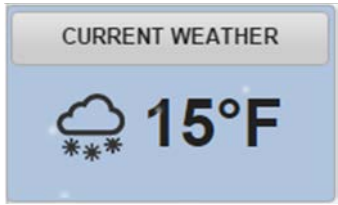 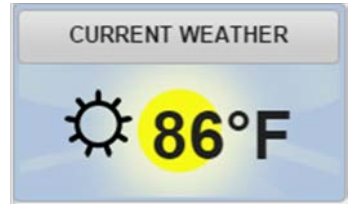

*Representing **winter***            *Representing **summer***

**Figure 2:** The use of weather widget to create hypothetical season.

**Figure 3:** Sample choice experiment task.

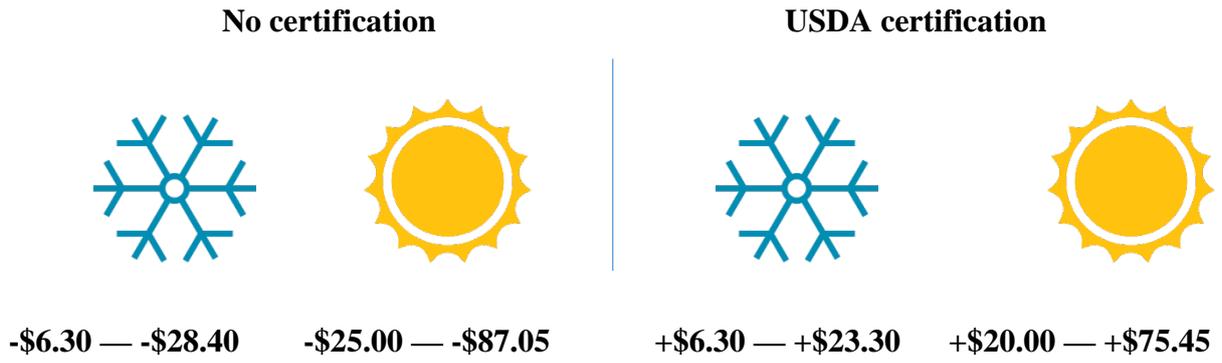

|  | No certification |  | USDA certification |  |
|---|---|---|---|---|
|  | ❄ | ☀ | ❄ | ☀ |
|  | -$6.30 — -$28.40 | -$25.00 — -$87.05 | +$6.30 — +$23.30 | +$20.00 — +$75.45 |

**Figure 4:** Willingness to pay ranges for certifications in winter and summer.

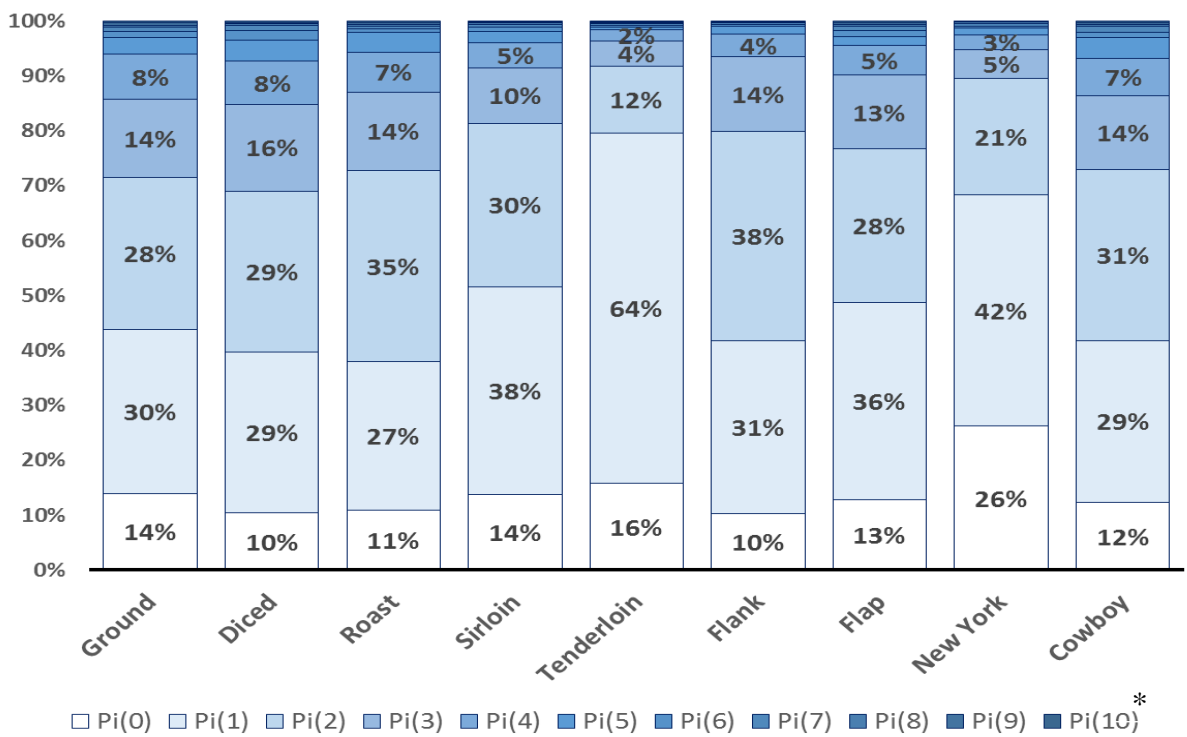

**Figure 5:** Probability of purchase quantity for each beef products in *winter*

*Pi represents the product probability of purchase and the number in the brackets presents the quantity of the product.

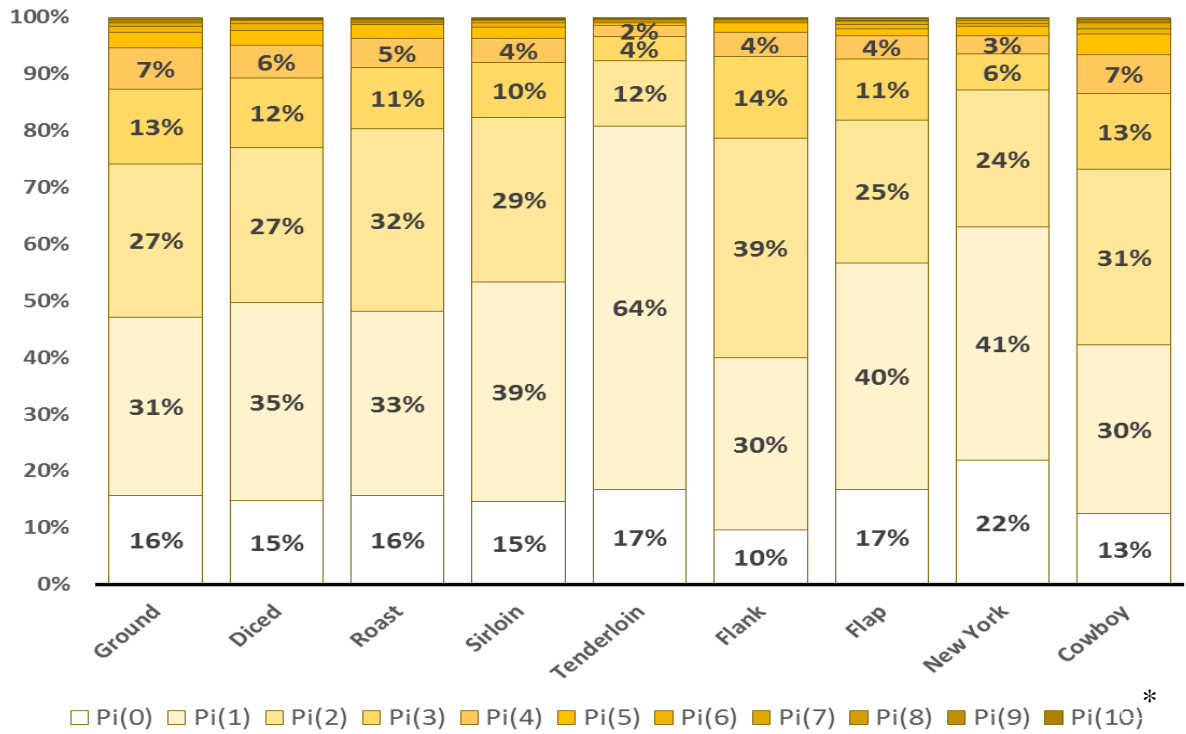

**Figure 6:** Probability of purchase quantity for each beef products in *summer*

*Pi represents the product probability of purchase and the number in the brackets presents the quantity of the product.